\documentclass[amsmath,amssymb,aps]{revtex4}

\usepackage{graphicx}
\usepackage{dcolumn}
\usepackage{bm}

\begin{document}


\title{Reconstruction of Residual Stress in a Welded Plate \\ 
Using the Variational Eigenstrain Approach}

\author{Alexander M. Korsunsky}
\email{alexander.korsunsky@eng.ox.ac.uk}
\affiliation{Department of Engineering Science, University of Oxford\\ 
Parks Road, Oxford OX1 3PJ, England}

\author{Gabriel M. Regino}
\affiliation{Department of Engineering Science, University of Oxford\\ 
Parks Road, Oxford OX1 3PJ, England}

\author{David Nowell}
\affiliation{Department of Engineering Science, University of Oxford\\ 
Parks Road, Oxford OX1 3PJ, England}

\date{\today}

\begin{abstract}

We present the formulation for finding the distribution of eigenstrains, i.e. the sources of residual stress, from a set of measurements of residual elastic strain (e.g. by diffraction), or residual stress, or stress redistribution, or distortion. The variational formulation employed seeks to achieve the best agreement between the model prediction and some measured parameters in the sense of a minimum of a functional given by a sum over the entire set of measurements. The advantage of this approach lies in its flexibility: different sets of measurements and information about different components of the stress-strain state can be incorporated. We demonstrate the power of the technique by analysing experimental data for welds in thin sheet of a nickel superalloy aerospace material. Very good agreement can be achieved between the prediction and the measurement results without the necessity of using iterative solution. In practice complete characterisation of residual stress states is often very difficult, due to limitations of facility access, measurement time or specimen dimensions. Implications of the new technique for experimental analysis are all the more significant, since it allows the reconstruction of the entire stress state from incomplete sets of data.

\end{abstract}

\maketitle

\section{\label{sec:intro}Introduction}

Reliable prediction of durability of structural components and assemblies is a fundamental requirement in various branches of engineering: transport, power generation, manufacturing and many others. This requirement led to the development of various analytical approaches to structural integrity, including elasto-plastic fracture mechanics, low cycle and high cycle fatigue analysis, creep and damage analysis, and to the introduction and standardization of appropriate methods of material property characterization. Once the material properties are determined from series of controlled laboratory experiments, numerical models of complex assemblies are used to predict the location and time of failure. Underlying this methodology is the assumption that material properties and damage accumulation models validated in the laboratory setting can be successfully transferred to the prototype (full scale object), provided certain requirements of scale independence are fulfilled. 

One of the difficulties arising in the way of applying this methodology is the fact that in the process of assembly materials undergo additional processing operations that modify their internal structure (e.g. grain size and texture, composition) and internal load distribution (residual stress). Residual stress appears to be a particularly difficult parameter to account for. Unlike e.g. microstructure and composition, residual stress is associated with the entire assembly, and often undergoes significant changes if a testing piece is extracted from it for investigation. Yet residual stress is often the most crucial parameter that determines the durability of an assembled structure.

Welding is a joining and fabrication technique that relies on local melting or softening of the parent material with the purpose of allowing it to fuse together with the filler material and the opposing piece to which a joint is being made. In the process of welding the material undergoes a complex thermal and mechanical history, and the resulting assembly inherits a memory of this process in the form of microstructural and compositional changes, and residual stress distribution. In the recent decades significant efforts have been devoted by many researchers to the development of detailed numerical models of the welding process; yet at the present time reliable prediction of material structure and residual stress state at the outcome of a welding process remains elusive, its use being primarily limited to qualitative identification of, for example, the most favourable conditions for producing a weld with lower residual stress and distortion.

However, it remains necessary to predict durability of assemblies in the presence of welding-induced residual stresses, since this joining method remains in widespread use e.g. in the aerospace industry. In this situation experimental methods of residual stress determination come to the fore, since they provide information about the central link in the chain {\em processing} - {\em residual stress} - {\em structural integrity}. Methods of residual stress evaluation can be notionally split into {\em mechanical} and {\em physical}. Mechanical methods of residual stress determination rely on detecting the deformation (strain) or distortion in the test piece that arises following some cutting or material removal. Perhaps the most well-known of such techniques is hole drilling, that involves plunging a sharp fast drill into the surface of material, and detecting strain change from the original state at the surface of the material using a specially designed strain rosette. The results of a hole drilling experiment are interpreted using an elastic numerical model. Another method developed recently is known as the contour method \cite{prime} in which the test piece is carefully sliced using spark erosion and a two-dimensional map of displacement in the direction normal to the cutting plane is collected. This map is then used as the input for an elastic finite element numerical model of the piece, and allows the residual stress to be determined in the plane of the section.

Physical methods of residual stress evaluation rely on the determination of some parameter of the system that is known to correlate with the residual stress. Perhaps the most well-known of the physical methods of residual stress determination is X-ray diffraction. In a diffraction experiment a beam of particles (e.g. photons or neutrons) is directed at a particular location within a polycrystalline sample, and an intensity profile of the scattered particles is collected, either as a function of scattering angle for fixed energy beam (monochromatic mode), or as a function of photon energy for fixed scattering angle (white beam mode). In both cases the pattern displays distinct peaks that correspond to distances between crystal lattice planes that are prevalent within the sample. If strain-free distances are known for the same sets of planes, then the measurements allow the calculation of residual direct elastic strains referring to specific orientations within the crystal and the sample. 

The most widespread laboratory implementation of the X-ray diffraction method for the determination of residual stress is known as the $\sin^2\psi$ technique. In this technique a series of measurements is carried out to collect the data for elastic lattice strains for a set of directions that deviate from the sample normal by a varying angle $\psi$. An assumption is then made that these measured strains correspond to a consistent strain state within homogeneous isotropic linear elastic solid, that allows the stress state within the sample to be deduced. An important observation that needs to be made at this point concerns the fact that residual stress is not {\em measured} in such an experiment, but merely deduced on the basis of certain assumptions, including that (i) that the material is uniform, isotropic and continuous, (ii) that strain values measured at different angles of tilt, $\psi$, are obtained from the same group of grains within the same gauge volume; (iii) that the component of stress normal to the sample surface vanishes within the gauge volume; etc. The above assumptions are in fact approximations whose validity may or may not be readily proven, or which are, in the worst case, simply wrong. 

The diffraction of neutrons and high energy synchrotron X-rays provides a unique non-destructive probe for obtaining information on strains deep in the bulk of engineering components and structures, e.g. \cite{JSR}. This method has become a mature tool for the determination of residual strain states in small coupons, and developments are under way to establish the facilities for performing high resolution measurements directly on larger engineering components \cite{epsrc}.

A particular feature of high energy X-ray diffraction is that the radiation is primarily scattered forward, i.e. in directions close to that of the incident beam \cite{liu}. Therefore small diffraction angles have to be used, usually $2\theta<15^\circ$. Two difficulties follow. Firstly, it is difficult to measure strains in directions close to that of the incident beam. This is due to the fact that the scattering vector is always the bisector of the incident and diffracted beams. Hence for high energy X-rays the strain measurement directions form a shallow cone. For a scattering angle of $2\theta$  this cone has the angle of $(180^\circ-2\theta)/2=90^\circ-\theta$. In practice this means that strain directions accessible for the high energy X-ray diffraction techniques are close to being normal to the incident beam. This situation is in stark contrast with that encountered in laboratory X-ray diffraction where near backscattering geometry is used, and measured strains are in directions close to being parallel with the incident beam. Secondly, it is difficult to achieve good spatial resolution in the direction of the incident beam, due to the elongated shape of the scattering volume. Although rotating the sample may help to overcome these difficulties, in practice this is often limited by the sample shape and absorption, and means that often only two components of strain (in the sample system) are known with sufficient accuracy and resolution.

Neutron diffraction strain analysis has the advantage that it is more often possible to measure the lattice parameter in three mutually perpendicular directions. It is therefore sometimes claimed that this is the only method capable of 'true' residual stress measurement. However, it must be noted even then that the residual stress evaluation involves making certain assumptions: that indeed three principal directions were chosen; that the strain-free lattice parameters have been correctly determined for all three directions; that the correct values of Young's modulus and Poisson's ratio were used in the calculation. In other words, stress evaluation relies on calculations based on certain assumptions. Furthermore, the conventional interpretation procedures remain point-wise, i.e. make no attempt to establish correlation between measurements at different locations, and to check whether the results conform to the basic requirements of stress and moment balance within each arbitrarily chosen sub-volume, and that strain compatibility and traction-free surface conditions are satisfied. 

The purpose of the foregoing discussion was to establish the basic fact that residual stress state is never measured directly, be it by mechanical or physical methods, but always deduced by manipulating the results of the measurements in conjunction with certain models of deformation. 

It is therefore correct to say that residual stress measurement is one area of experimental activity where the development and validation of numerical models needed for the interpretation of data occupies a particularly important place: without adopting some particular model of deformation is it impossible to present measurement results in terms of residual stress.

To give a very general example, when a ruler is pressed against the sample to determine its length, the implication is that the sample and ruler surfaces are collocated all along the measured length; and that the length of the ruler between every pair of markers is exactly the same. Only if that is so then the reading from the ruler is correct. 

The approach adopted in this study rests on the explicit postulate that it is necessary to make informed assumptions about the nature of the object (or distribution) that is being determined in order to regularise the problem. Interpretation of any and every measurement result relies on a model of the object being studied.

In the present paper we are concerned with a model of residual stress generation to which we refer as the eigenstrain technique. The term eigenstrain and notation $\epsilon^*$ for it were introduced by Toshio Mura \cite{mura} to designate any kind of permanent strain in the material arises due to some inelastic process such as plastic deformation, crystallographic transformation, thermal expansion mismatch between different parts of an assembly, etc. In some simple cases, e.g. in the analysis of residually stressed bent beams, it is possible to derive explicit analytical solutions for residual stresses as a function of an arbitrary eigenstrain distribution \cite{JSA}. In the more general context it is apparent that eigenstrains are responsible for the observed residual stresses, and therefore can be thought of as the source of residual stress; yet eigenstrain is not a priori associated with any stress, nor can it be measured directly, as it does not manifest itself in the change of crystal lattice spacing or any other physical parameter. In fact, if a uniform eigenstrain distribution is introduced into any arbitrarily shaped object, then no residual stress is produced. In other words, eigenstrains are associated with misfit and mismatch between different parts of an object. This conclusion is particularly interesting on the context of the foregoing discussion about engineering assemblies and the nature of residual stresses in them. 

The following discussion is based on the analysis of the fundamental equations describing the generation of residual elastic stresses and strains from a distribution of eigenstrains. Most often the problem has to be addressed in the inverse sense: residual stresses or residual elastic strains are somehow known in a number of locations, while the unknown underlying distribution of eigenstrain sources of residual stresses needs to be found. While the direct problem of residual stress determination from eigenstrain distribution can be classed as easy (linear, elastic), the inverse problem is more difficult. However, it is important to emphasize that once the eigenstrain distribution is found, it can be used to solve the 'easy' direct problem, so that the residual stress distribution becomes known in full. 

The procedure of finding the underlying eigenstrain distribution and reconstructing the complete residual stress state is entirely similar in principle to any other method of residual stress determination discussed above: the experimental data are treated using some suitable numerical model, and the residual stress state is deduced. There are some distinct advantages offered by the eigenstrain approach. Firstly, the solution obtained in this way is forced to satisfy the requirements of total strain compatibility and stress equilibrium, that often become violated if less sophisticated methods of data interpretation are used. Secondly, once the eigenstrain distribution is deduced it allows the prediction of the object's distortion and residual stress re-distribution during any subsequent machining operation, such as sectioning or surface layer removal.

In the following section we present a formulation of the direct problem of residual stress determination from the known eigenstrain distribution. We then formulate an efficient non-iterative variational approach for solving the inverse problem, and describe briefly its possible numerical implementations. We then apply the method to the analysis of experimental data for residual elastic strains in a single pass electron beam weld in a plate of aerospace nickel superalloy IN718, obtained using high energy synchrotron X-ray diffraction. We demonstrate how the eigenstrain distribution can be found  that minimizes the disagreement between the measurements and the model prediction, and also how the method allows the refinement of the strain-free lattice parameter across the weld. We show reconstructions of the complete residual stress field within the plate. Finally, we carry out sensitivity analysis to determine whether the solution obtained in terms of the eigenstrain distribution (and hence in terms of the reconstructed residual stress state) is stable with respect to the amount of experimental residual elastic strain data available. 

\section{\label{sec:exp} Experimental}

\begin{figure}
\centerline{ \includegraphics[width=10.cm]{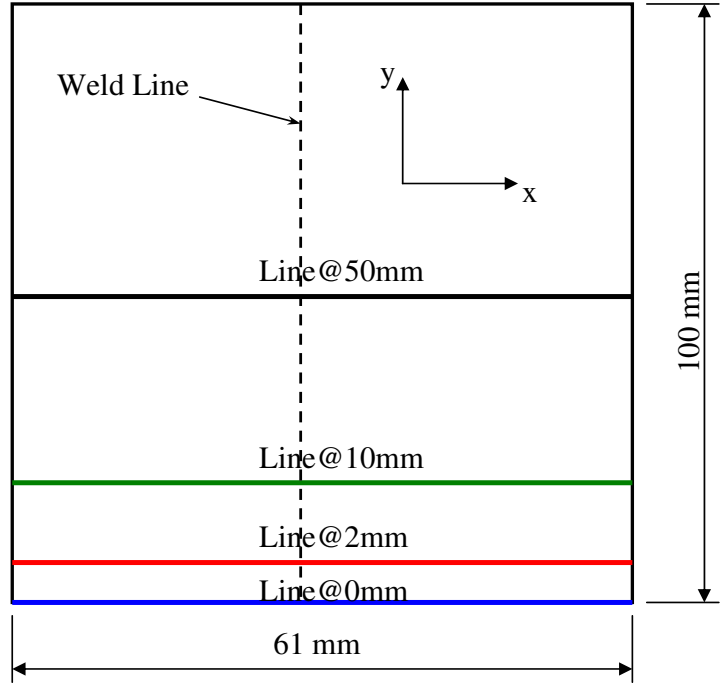} }
\caption{
Geometry of the lower half of the welded plate of nickel superalloy IN718 considered in the present study. Synchrotron X-ray diffraction measurements of strains in the longitudinal direction, $\epsilon_{yy}$, and transverse direction, $\epsilon_{xx}$, were carried out along each line, allowing macroscopic residual stresses to be calculated. 
}
\label{fig:one}
\end{figure}

Figure \ref{fig:one} illustrates the dimensions of the experimental specimen used in the present study, and the location of the measurement points. Electron beam welding was used to manufacture a flat rectangular plate by joining two elongated strips (3mm thick, 200mm long and approximately 25 and 35mm wide).

The sample used for the experiment was made from IN718 creep resistant nickel superalloy used in the manufacture of aeroengine components, such as combustion casings and liners, as well as disks and blades. The composition of the alloy in weight percent is approximately given by 53\% Ni, 19\% Fe, 18\% Cr, 5\% Nb, and small amounts of additional alloying elements Ti, Mo, Co, and Al. Apart from the matrix phase, referred to as $\gamma$, the microstructure of the alloy may show additional precipitate phases, referred to as $\gamma'$, $\gamma''$, and $\delta$.
 
The primary strengthening phase, $\gamma''$, has the composition $\rm Ni_3 Nb$ and a body-centred tetragonal structure, and forms semi-coherently as disc-shaped platelets within the $\gamma$ matrix. It is highly stable at $600^\circ$C, but above this temperature it decomposes to form the $\gamma'$ $\rm Ni_3 Al$ phase (between $650^\circ$C and $850^\circ$C), and $\delta$, having the same composition as $\gamma''$ (between $750^\circ$C and $1000^\circ$C). At large volume fractions and when formed continuously along grain boundaries, the $\delta$ is thought to be detrimental to both strength and toughness \cite{Brooks}. The $\delta$ phase that forms is more stable than the $\gamma''$ phase, and has an orthorhombic structure \cite{Guest}.

Welding is known to give rise to residual tensile stresses along the weld line and in the immediately adjacent heat affected zones (HAZ), while compressive residual stress is found further away from the seam. The most significant residual stress component is the longitudinal stress that we denote by symbols $\sigma_{22}$ or $\sigma_{yy}$ that have the same meaning throughout.

High energy synchrotron X-ray diffraction measurements were carried out on the ID11 and ID31 beamlines at the ESRF using monochromatic mode and a scanning diffractometer. The energy of about 70keV was selected by the monochromator, and the 111 reflection of the $\gamma$ matrix phase was used. The beam spot size on the sample was approximately 1mm (horizontal) by 0.25mm (vertical). 

Line scans were performed with the scan step size of 1mm along the four lines indicated in Figure \ref{fig:one}, lying 0mm, 2mm, 10mm and 50mm above the lower edge of the weld plate. Both the horizontal (transverse) strain component, $\epsilon_{xx}$, and the vertical (longitudinal) strain component $\epsilon_{yy}$ were evaluated. Assuming that the state of plane stress persists in the thin welded plate, this allowed the stress components $\sigma_{xx}$ and $\sigma_{yy}$ to be calculated.

\begin{figure}
\centerline{ \includegraphics[width=12.cm]{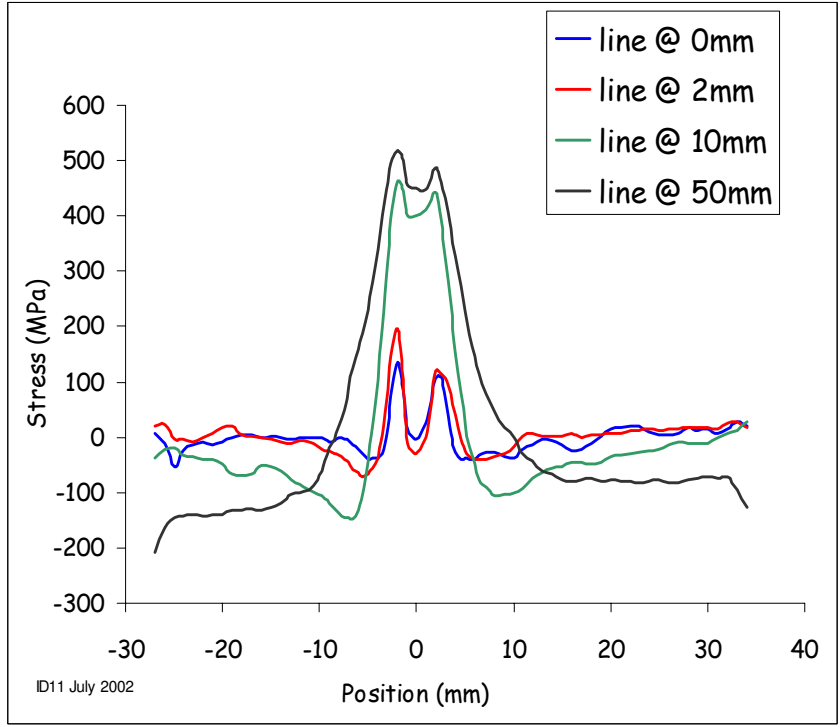} }
\caption{
Variation of the residual stress component in the longitudinal direction, $\sigma_{yy}$, along each of the four lines, calculated from two mutually orthogonal components of strain under the assumption of plane stress. Note that the curve corresponding to the edge of the plate (0mm) does not show uniform zero stress. 
}
\label{fig:two}
\end{figure}

Figure \ref{fig:two} illustrates the results of the measurement interpreted in this straightforward way, plotted as the longitudinal residual stress $\sigma_{yy}$ as a function of horizontal position measured from the nominal centre of the weld line. It is apparent from the plots that the stress profiles evolve both in terms of the magnitude and shape away from the edge of the plate.

Originally the results were interpreted by assuming a constant value of $d_0$, the unstrained lattice spacing, everywhere within the plate. However, this led to the physically unacceptable result of non-zero longitudinal stress existing at the bottom edge of the plate. The calculation was then repeated imposing the stress-free condition at the bottom edge, and allowing $d_0$ to vary only as the function of the horizontal coordinate $x$ along the bottom edge of the weld plate so as to produce stress free condition at that edge.

\section{\label{sec:vareig}Variational eigenstrain analysis}

Distributions of inelastic strains contained in the sample act as sources of residual stresses. Indeed, in the absence of inelastic (permanent) strain of some origin (or indeed, when such inelastic strain is uniform throughout the sample),  then any specimen is stress free in the absence of external loading. For a known non-uniform eigenstrain distribution $\epsilon_{ij}^*({\bf x}')$ the elastic residual strains (and hence residual stresses) in the body can be found by the formulae \cite{mura}
\begin{equation}
e_{kl}({\bf x}) = -\epsilon_{kl}^*({\bf x})-\int_{-\infty}^{\infty} C_{pqmn}
\epsilon_{mn}^*({\bf x}')G_{kp,ql}({\bf x},{\bf x}'){\rm d}{\bf x}', \quad
\sigma_{ij}({\bf x})=C_{ijkl} e_{kl}({\bf x}).
\nonumber
\end{equation}
The above formula is in principle applicable to bodies of arbitrarily complex shape, provided the elastic constants $C_{ijkl}$ are known, together with the corresponding Green's function $G_{kp}({\bf x},{\bf x}')$. In practice Green's functions can be found only for bodies of simple geometry, e.g. infinitely extended two-dimensional or three-dimensional solid. The fundamental value of the above formula, however, lies in the statement that for known eigenstrain distribution the elastic response of the body containing it can be readily found. 

For convoluted geometries the finite element method provides a suitable method of solving the above direct problem of finding residual elastic strains from given eigenstrains. We are interested here in the problem that often arises in residual stress measurement and interpretation. Let there be given a set of measurements (with certain accuracy) of strains and stresses collected from a finite number of points (sampling volumes) within a bounded specimen. We would like to solve the inverse problem about the determination of unknown eigenstrains  from this incomplete knowledge of elastic strains or residual stresses.  The limited accuracy and lack of completeness of measurements suggest that direct inversion of (1) may not be the preferred solution. In fact the method chosen must be sufficiently robust to furnish approximate solutions even in this case.

The incorporation of eigenstrain into the finite element method framework can be accomplished via the use of anisotropic pseudo-thermal strains. In the present case we concentrated our attention on the determination of a single eigenstrain component, $\epsilon_{22}*$, longitudinal with respect to the extent of the weld joint. It is clear that in practice this is not the only eigenstrain component likely to be present. However, it is also apparent that this component is the most significant in terms of its effect on the longitudinal stress, $\sigma_{22}$. It is worth noting that the procedure outlined below is not in any way restricted to the determination of single component eigenstrain distributions, but is in fact entirely general.

\begin{figure}
\centerline{ \includegraphics[width=12.cm]{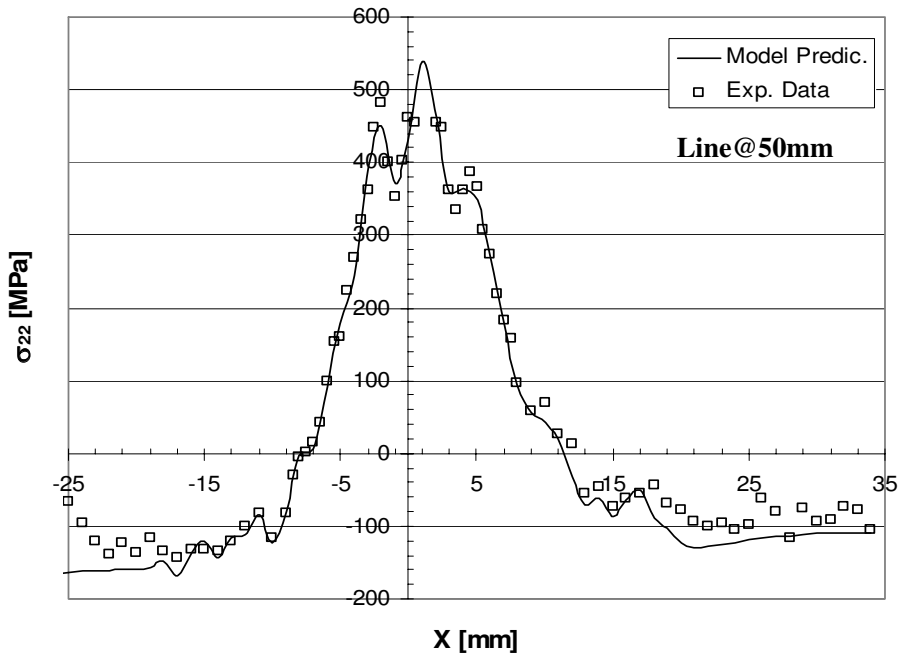} }
\caption{
Prediction of the variational eigenstrain model using the complete data set available for the residual stress component $\sigma_{22}=\sigma_{yy}$ along the line 50mm above the lower edge of the welded plate. Stresses computed from measurements are shown as markers, while the model prediction is shown by the continuous curve. The stress values are plotted as a function of horizontal position $x=x_1$ with respect to the nominal centre of the weld line.
}
\label{fig:three}
\end{figure}

The eigenstrain distribution was introduced in the form of a truncated series of basis functions
\begin{equation}
\epsilon^*(x,y)=\sum_{k=1}^K c_k E_k(x,y).
\label{eq:one}
\end{equation}
Each of the basis functions $E_k(x,y)$ is chosen in the variable separable form as
\begin{equation}
E_k(x,y)=f_i(x) g_j(y),
\end{equation}
and index $k$ provides numeration for the entire pair set $(i,j)$. Functions $f_i(x)$ and $g_j(y)$ are chosen to reflect the nature of the eigenstrain dsitribution in the present problem. It was found that for the functions of the horizontal coordinate, $f_i(x)$, a suitable choice is provided by the Chebyshev polynomials $T_i(\overline{x}), \quad i=0..I$, with the argument $\overline(x)$ scaled from the canonical interval $[-1,1]$ to the plate width. For the functions of the vertical coordinate, $g_j(y)$, powers $y^j, \quad j=0..J$, were used. 

As stated earlier, the solution of the direct eigenstrain problem can be readily obtained for any eigenstrain distribution by an essentially elastic calculation within the FE model. This task is easily accomplished for the basis functions $E_k(x,y)=f_i(x) g_j(y)$. Furthermore, due to the problem's linearity, the solution of the direct problem described by a linear combination of individual eigenstrain basis functions $E_k(x,y)=f_i(x) g_j(y)$ with coefficients $c_k$ is given by the linear superposition of solutions with the same coefficients. This observation provides a basis for formulating an efficient variational procedure for solving the inverse problem about the determination of underlying eigenstrain distribution. 

The problem that we wish to address here stands in an inverse relationship to the direct eigenstrain problem. This is the situation most commonly encountered in practice: the residual elastic strain distribution is known, at least partially, e.g. from diffraction measurement. The details of the preceding deformation process need to be found, such as distribution of eigenstrains within the plastic zone. Alternatively, in the absence of non-destructive measurements of residual elastic strain, changes in the elastic strain may have been monitored, e.g. using strain gauges, in the course of material removal. 

Questions arise immediately regarding the invertibility of the problem; its uniqueness; the regularity of solution, i.e. whether the solution depends smoothly on the unknown parameters. Although do not attempt to answer these questions here, we present a constructive inversion procedure and also provide a practical illustration of its stability.

Denote by $s_k(x,y)$ the distribution of the longitudinal stress component $\sigma_{yy}$ arising from the eigenstrain distribution given by the $k-$th basis function $E_k(x,y)$. Evaluating $s_k(x,y)$ at each of the $q-$th measurement points with coordinates $(x_q,y_q)$ gives rise to predicted values $s_{kq}=s_k(x_q,y_q)$. Due to the problem's linearity the linear combination of basis functions expressed by equation (\ref{eq:one}) gives rise to the stress values at each measurement point given by the linear combination of $s_{kq}$ with the same coefficients $c_k$, i.e. $\sum_k c_k s_{kq}$.

Denote by $t_q$ the values of the same stress component $\sigma_yy$ at point $(x_q,y_q)$ deduced from the experiment. In order to measure the goodness of the prediction we form a functional $J$ given by the sum of squares of differences between actual measurements and the predicted values, with weights:
\begin{equation}
J=\sum_q w_q \left( \sum_k c_k s_{kq}-t_q \right)^2,
\label{eq:three}
\end{equation}
where the sum in $q$ is taken over all the measurement points. The choice of weights $w_q$ remains at our disposal and can be made e.g. on the basis of the accuracy of measurement at different points. In the sequel we assume for simplicity that $w_q=1$, although this assumption is not restrictive. 

The search for the best choice of model can now be accomplished by minimising $J$ with respect to the unknown coefficients, $c_k$, i.e. by solving
\begin{equation}
{\rm grad}_{c_k} J=(\partial J/\partial c_k)=0, \quad k=1..K. 
\label{eq:four}
\end{equation}

Due to the positive definiteness of the quadratic form (\ref{eq:three}), the system of linear equations (\ref{eq:four}) always has a unique solution that corresponds to a minimum of $J$.  

The partial derivative of $J$ with respect to the coefficient $c_k$  can be written explicitly as
\begin{equation}
\partial J/\partial c_k = 2 \sum_{q=1}^Q s_{kq} \left( 
\sum_{m=1}^K c_m s_{mq} - t_q \right)
= 2\left( \sum_{m=1}^K c_m \sum_{q=1}^Q s_{kq} s_{mq} - \sum_{q=1}^Q 
s_{kq} t_q \right) = 0.
\label{eq:five}
\end{equation}

We introduce the following matrix and vector notation
\begin{equation}
{\bf S} = \{ s_{kq} \}, \quad {\bf t}=\{t_q \}, \quad {\bf c}=\{ c_k\}.
\label{eq:six}
\end{equation}
The entities appearing in equation (\ref{eq:six}) can be written in matrix form as:
\begin{equation}
{\bf A} = \sum_{q=1}^Q s_{kq} s_{mq} = {\bf S\, S}^T, \quad 
{\bf b}=\sum_{q=1}^Q s_{kq} t_q = {\bf S\, t}.
\label{eq:seven}
\end{equation}
Hence equation (\ref{eq:five}) assumes the form
\begin{equation}
\nabla_{\bf c} J=2({\bf A\, c}-{\bf b})=0.
\label{eq:tri}
\end{equation}
The solution of the inverse problem has thus been reduced to the solution of the linear system
\begin{equation}
\bf A\, c = b
\label{eq:linsys}
\end{equation}
for the unknown vector of coefficients ${\bf c}=\{c_k\}$.

\begin{figure}
\centerline{ \includegraphics[width=12.cm]{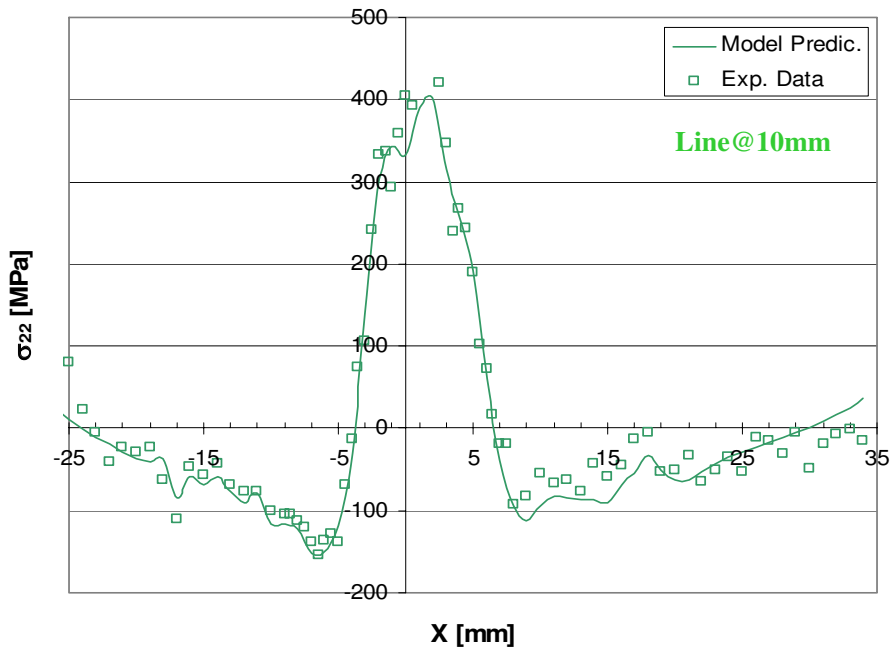} }
\caption{
Prediction of the variational eigenstrain model using the complete data set available for the residual stress component $\sigma_{22}=\sigma_{yy}$ along the line 10mm above the lower edge of the welded plate. Stresses computed from measurements are shown as markers, while the model prediction is shown by the continuous curve. The stress values are plotted as a function of horizontal position $x=x_1$ with respect to the nominal centre of the weld line.
}
\label{fig:four}
\end{figure}

\section{\label{sec:recon} Reconstructed stress fields}

Once the coefficients $c_k$ have been determined the eigenstrain distribution in equation (\ref{eq:one}) is introduced into the finite element model, and the complete stress-strain field is reconstructed by solving the direct problem. By construction the corresponding stress field satisfies the conditions of equilibrium within arbitrary sub-volume of the model, and traction-free boundary conditions are enforced. The total strain field composed of the residual elastic strains and eigenstrains satisfies the conditions of compatibility. The optimal agreement with the experimental measurements is achieved in the least squares sense over all eigenstrain distributions spanned by the functional space of equation (\ref{eq:one}).

\begin{figure}
\centerline{ \includegraphics[width=12.cm]{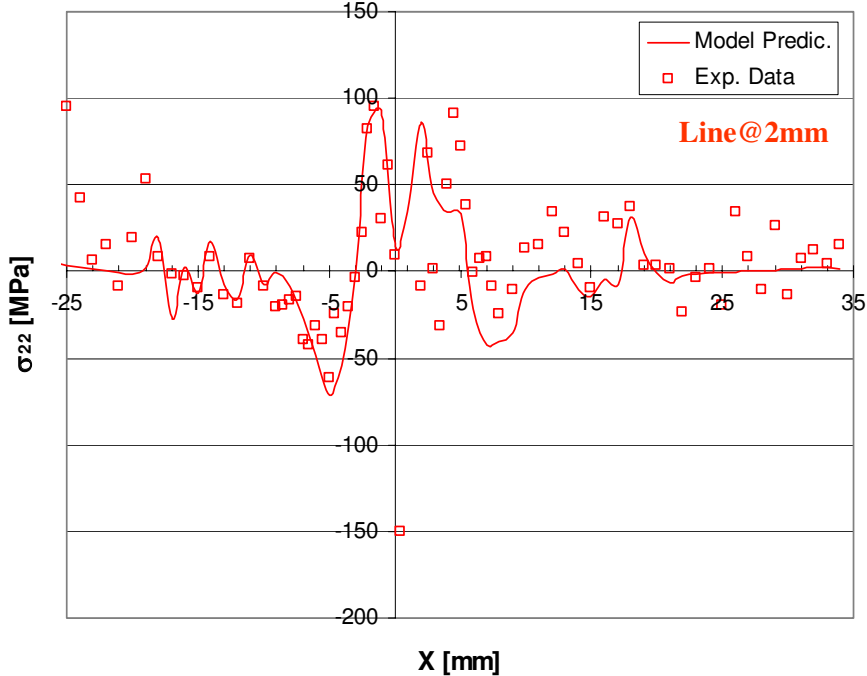} }
\caption{
Prediction of the variational eigenstrain model using the complete data set available for the residual stress component $\sigma_{22}=\sigma_{yy}$ along the line 2mm above the lower edge of the welded plate. Stresses computed from measurements are shown as markers, while the model prediction is shown by the continuous curve. The stress values are plotted as a function of horizontal position $x=x_1$ with respect to the nominal centre of the weld line.
}
\label{fig:five}
\end{figure}

Figure \ref{fig:three} shows the comparison between the experimental values shown by the markers and the reconstructed stress profile (continuous curve) along the line 50mm above the lower edge of the weld plate. Figure \ref{fig:four} shows a similar comparison for the line at 10mm from the edge, and Figure \ref{fig:five} for the line 2mm above the lower edge of the plate. Note the difference in the scales used for the three figures that explains the greater apparent scatter in the last plot. It is also worth recalling that as a result of adjustment of the $d_0$ values longitudinal stress $\sigma_{yy}$ is equal to zero along the line 0mm lying at the edge, and hence the plot is not shown. 

\begin{figure}
\centerline{ \includegraphics[width=12.cm]{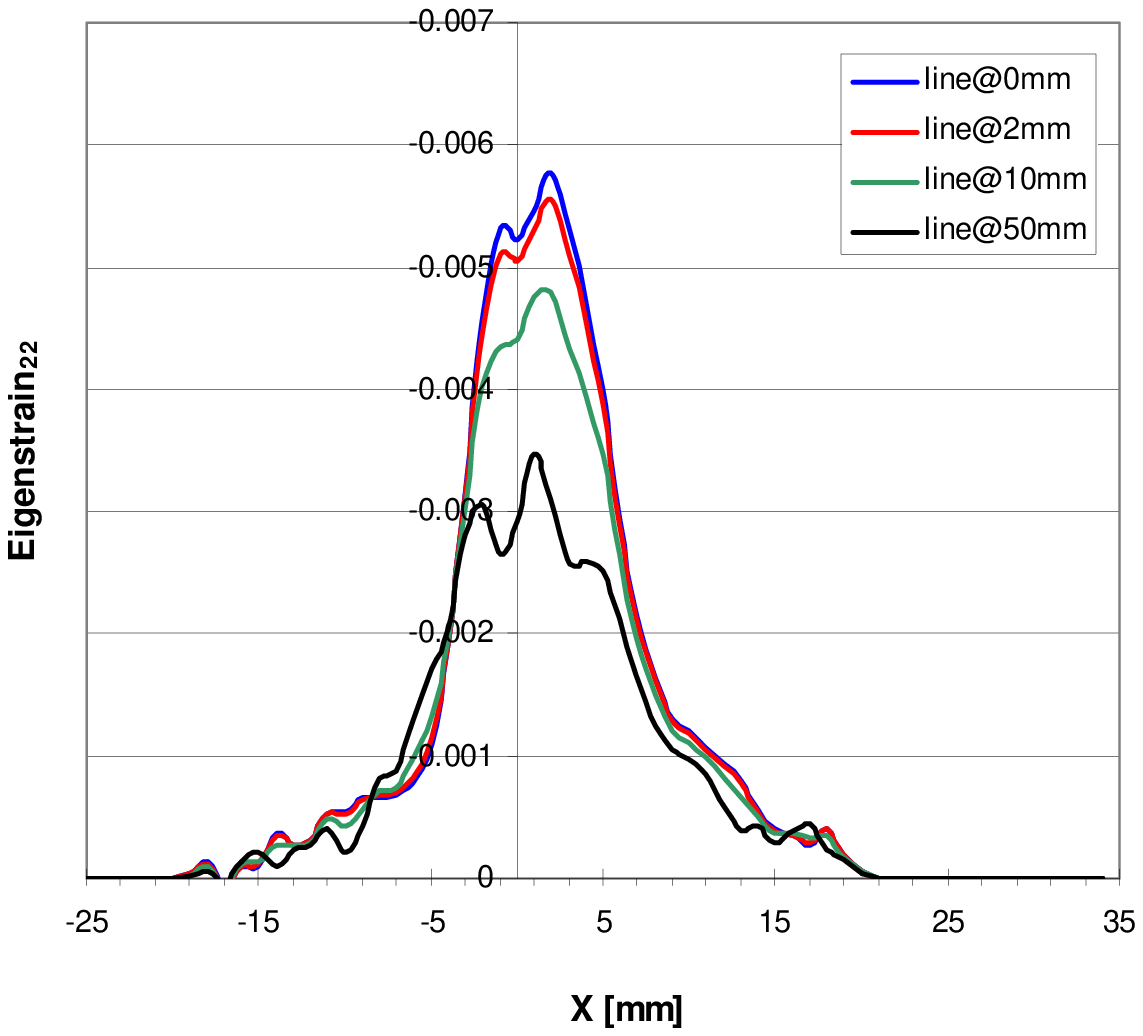} }
\caption{
An illustration of the nature of the eigenstrain distribution used in the variation model for residual stress reconstruction. The four curves indicate the variation of (compressive) eigenstrain as a function of the horizontal coordinate $x=x_1$ along the four lines lying 0mm, 2mm, 10mm and 50mm above the lower edge of the plate, respectively.
}
\label{fig:six}
\end{figure}

Figure \ref{fig:six} illustrates the nature of the eigenstrain distribution used in the variation model for residual stress reconstruction. The four curves indicate the variation of (compressive) eigenstrain as a function of the horizontal coordinate $x=x_1$ along the four lines lying 0mm, 2mm, 10mm and 50mm above the lower edge of the plate, respectively.

\begin{figure}
\centerline{ \includegraphics[width=12.cm]{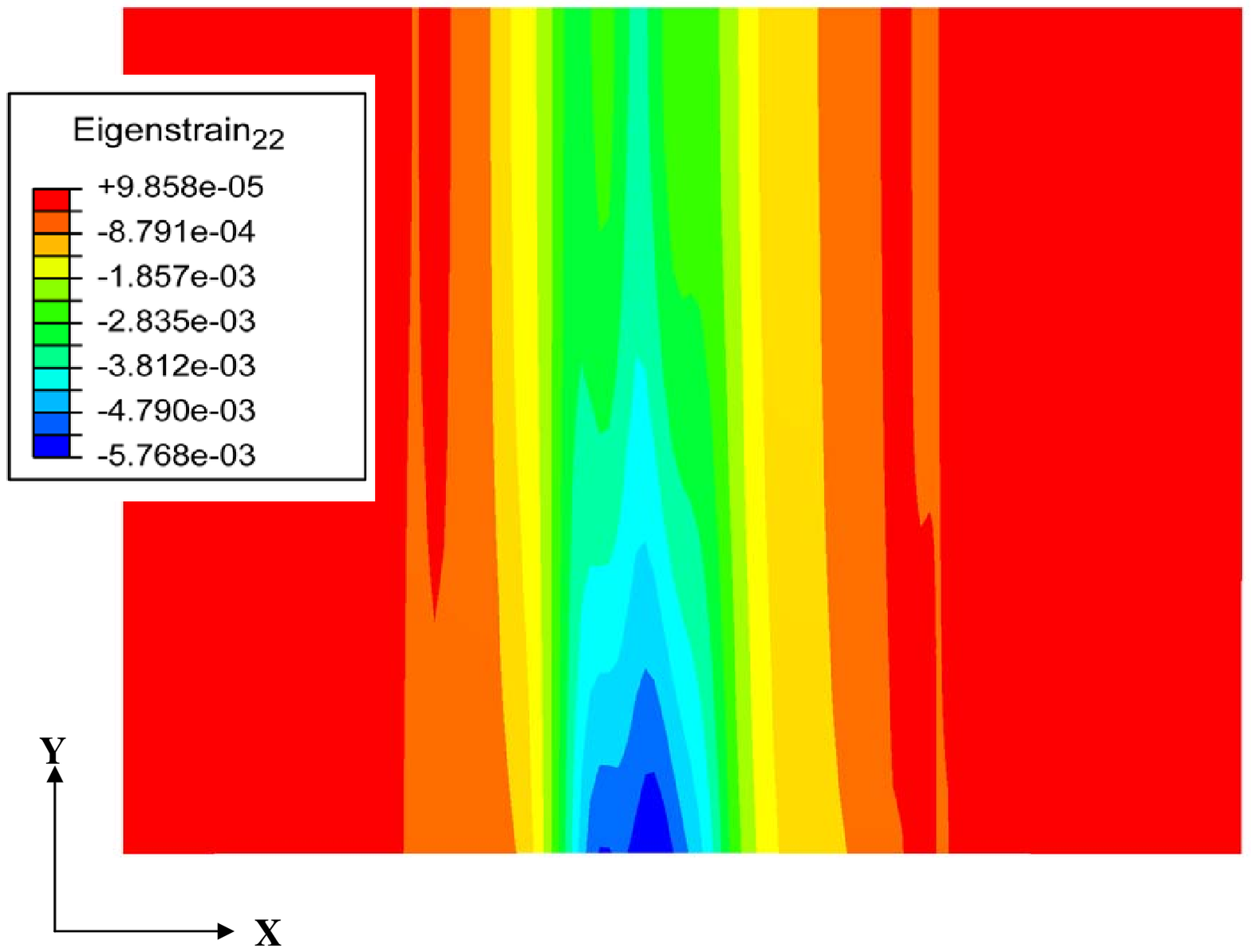} }
\caption{
The two-dimensional contour representation of the underlying eigenstrain field determined using the variational approach, shown for the lower half of the welded plate (symmetry is implied with respect to the upper edge of the plot).
}
\label{fig:seven}
\end{figure}

Figure \ref{fig:seven} shows a two-dimensional contour representation of the underlying eigenstrain field determined using the variational approach, shown for the lower half of the welded plate. Recall that symmetry is implied with respect to the upper edge of the plot.

\begin{figure}
\centerline{ \includegraphics[width=12.cm]{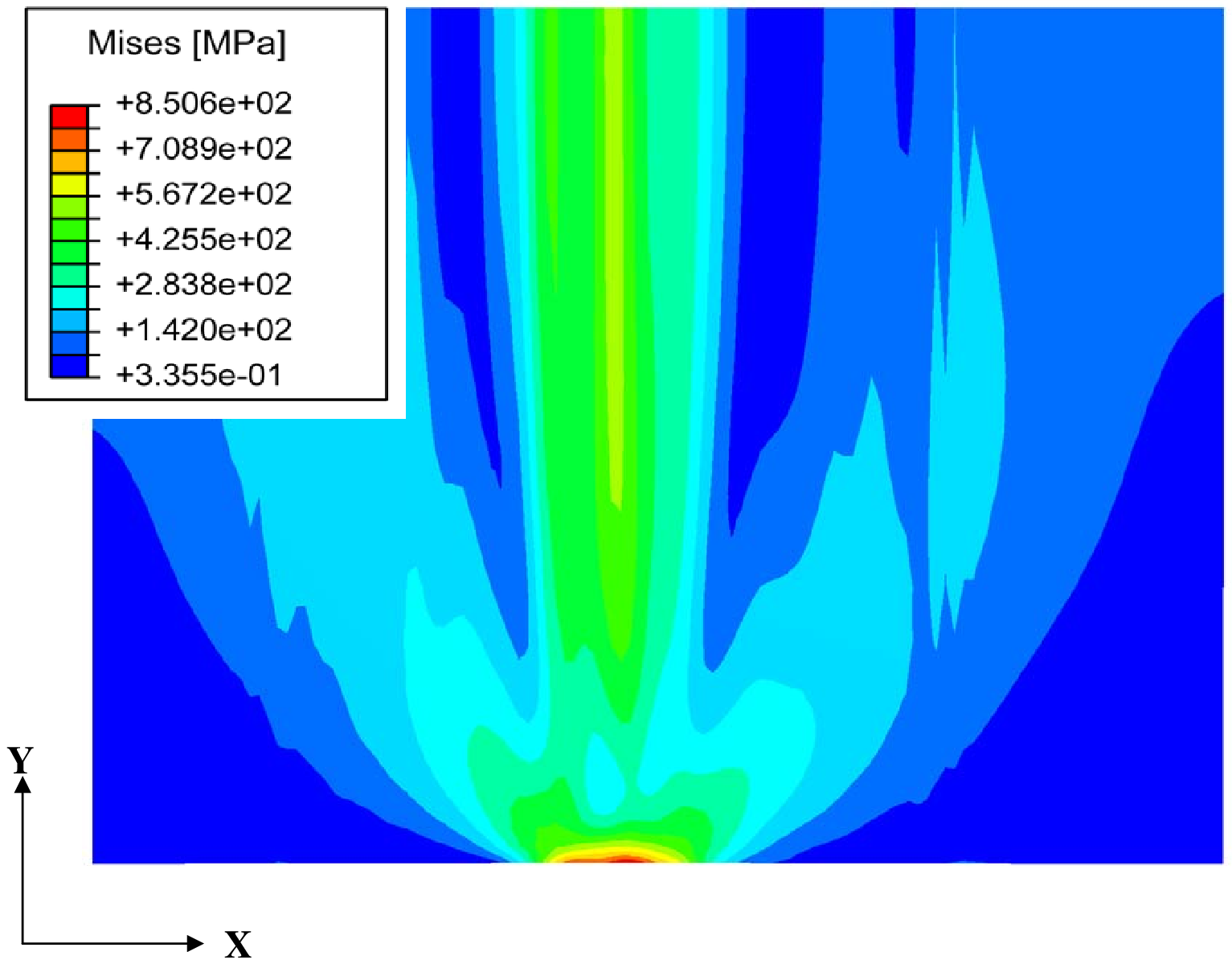} }
\caption{
Contour plot of the reconstructed von Mises stress field in the lower half of the welded plate.
}
\label{fig:eight}
\end{figure}

Figure \ref{fig:eight} shows a contour plot of the reconstructed von Mises stress field in the lower half of the welded plate.

\begin{figure}
\centerline{ \includegraphics[width=12.cm]{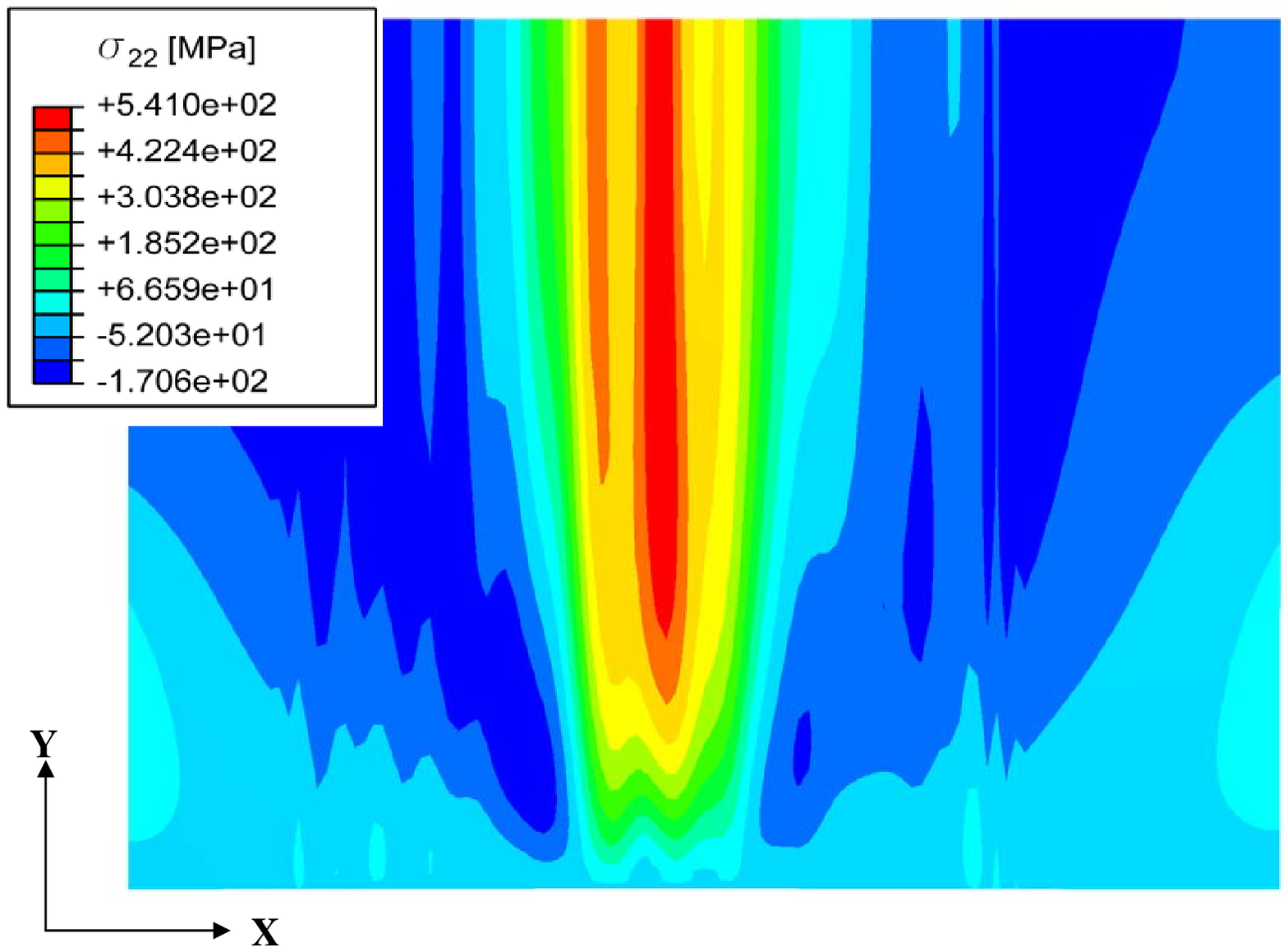} }
\caption{
Contour plot of the reconstructed vertical (i.e. longitudinal) stress component $\sigma_{yy}=\sigma_{22}$ in the lower half of the welded plate.
}
\label{fig:ten}
\end{figure}

Figure \ref{fig:ten} shows a contour plot of the reconstructed longitudinal $\sigma_{yy}$ stress field in the lower half of the welded plate.

\begin{figure}
\centerline{ \includegraphics[width=12.cm]{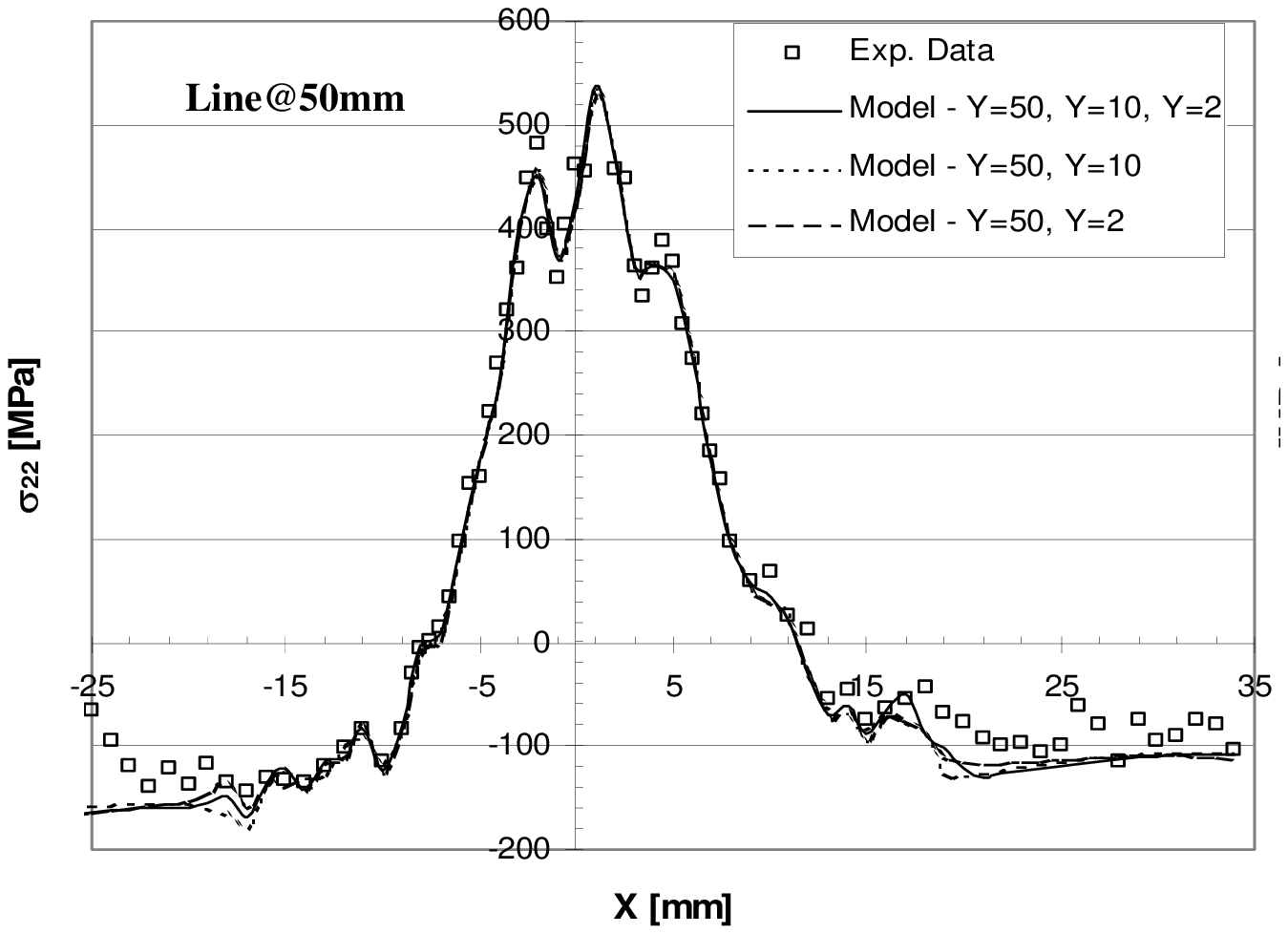} }
\caption{
Comparison plot for the reconstructed stress component $\sigma_{yy}=\sigma_{22}$ along the line at $y={\rm 50mm}$ from the lower edge of the plate from three models using the data for 50,10,2mm; 50,10mm; 50,2mm, respectively. 
}
\label{fig:twelve}
\end{figure}

\begin{figure}
\centerline{ \includegraphics[width=12.cm]{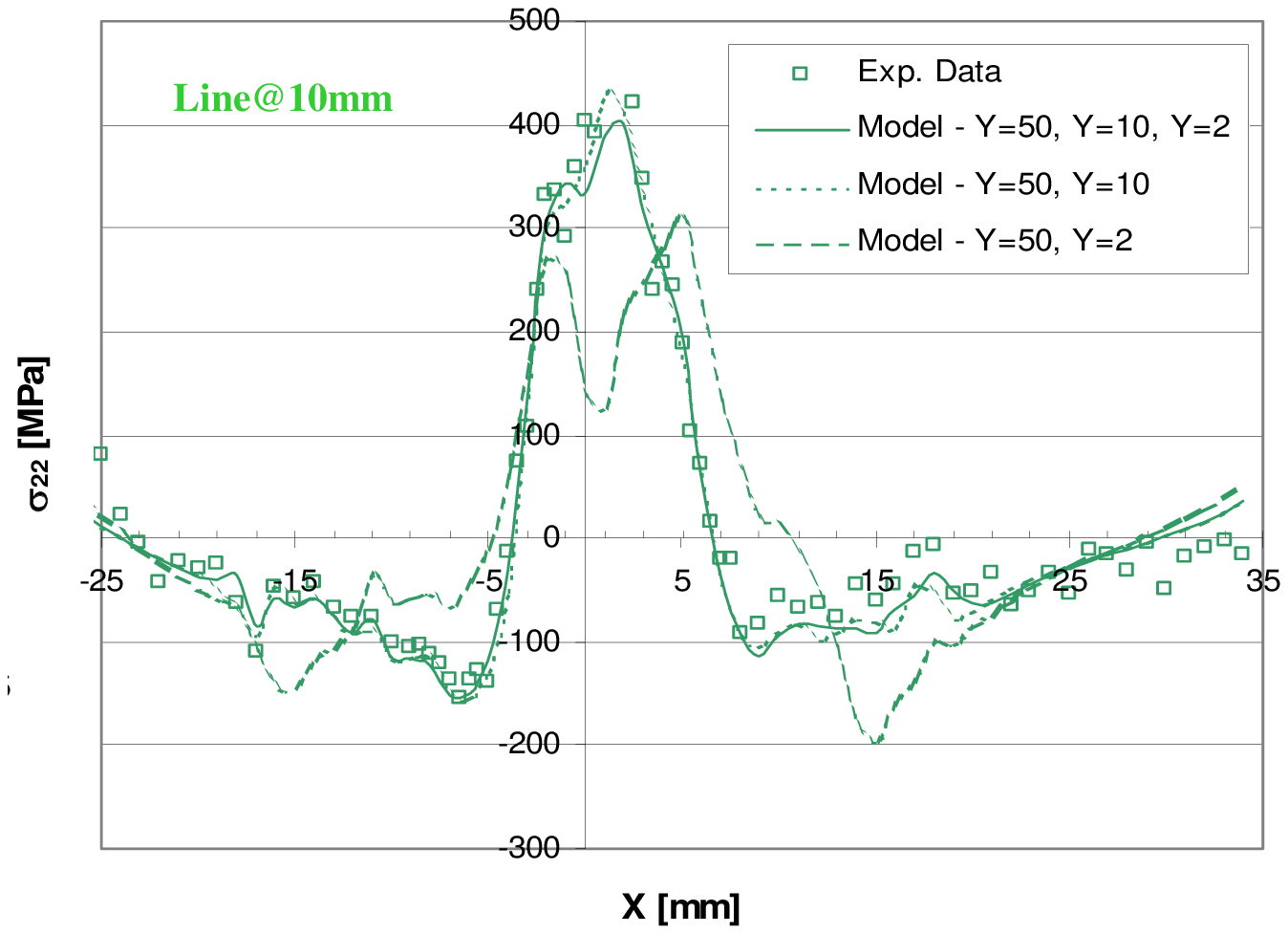} }
\caption{
Comparison plot for the reconstructed stress component $\sigma_{yy}=\sigma_{22}$ along the line at $y={\rm 10mm}$ from the lower edge of the plate from three models using the data for 50,10,2mm; 50,10mm; 50,2mm, respectively. 
}
\label{fig:thirteen}
\end{figure}

\begin{figure}
\centerline{ \includegraphics[width=12.cm]{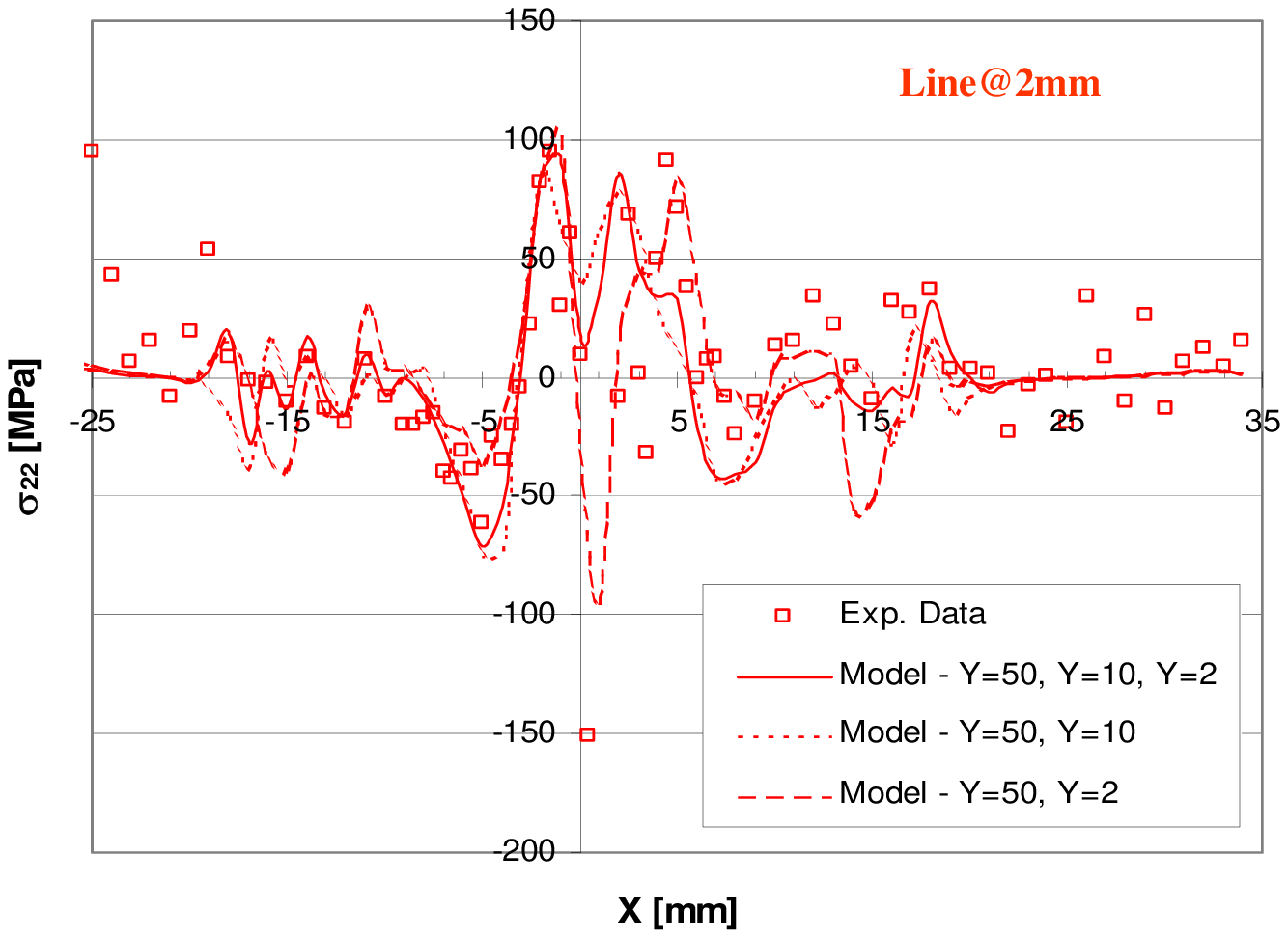} }
\caption{
Comparison plot for the reconstructed stress component $\sigma_{yy}=\sigma_{22}$ along the line at $y={\rm 2mm}$ from the lower edge of the plate from three models using the data for 50,10,2mm; 50,10mm; 50,2mm, respectively. 
}
\label{fig:fourteen}
\end{figure}

\begin{figure}
\centerline{ \includegraphics[width=12.cm]{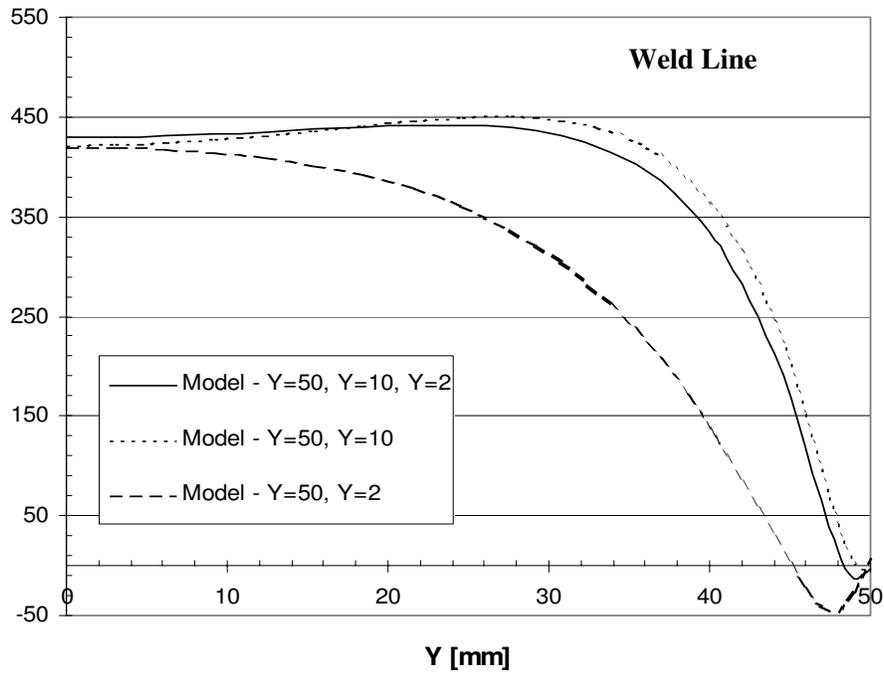} }
\caption{
Comparison plot for the reconstructed stress component $\sigma_{yy}=\sigma_{22}$ along the weld line $x=0$ obtained from three models using the data for 50,10,2mm; 50,10mm; 50,2mm, respectively. 
}
\label{fig:fifteen}
\end{figure}

\begin{figure}
\centerline{ \includegraphics[width=12.cm]{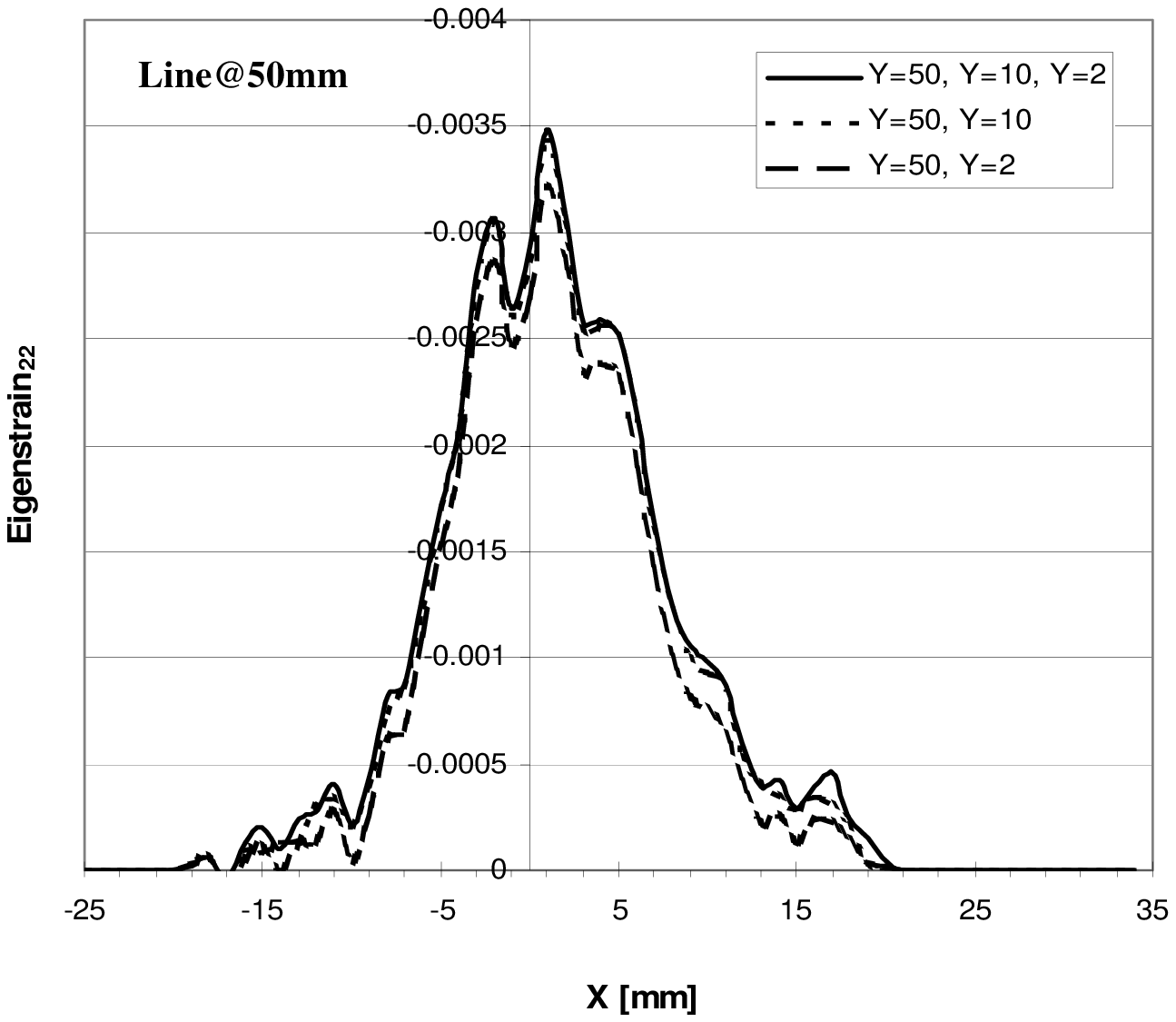} }
\caption{
Comparison plot for the eigenstrain variation perpendicular to the weld line at $y={\rm 50mm}$ obtained from three models using the data for 50,10,2mm; 50,10mm; 50,2mm, respectively. 
}
\label{fig:sixteen}
\end{figure}

\begin{figure}
\centerline{ \includegraphics[width=12.cm]{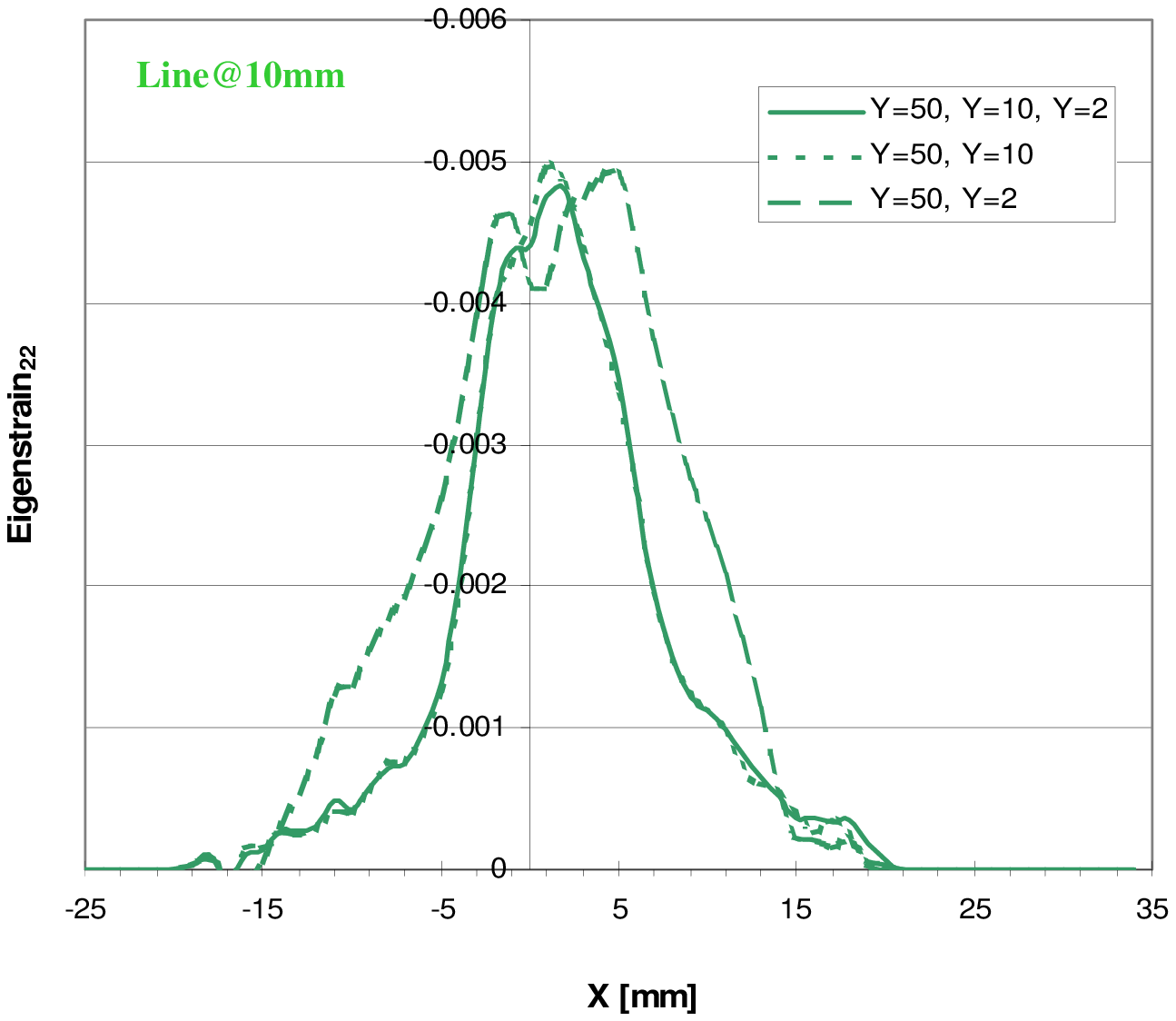} }
\caption{
Comparison plot for the eigenstrain variation perpendicular to the weld line at $y={\rm 10mm}$ obtained from three models using the data for 50,10,2mm; 50,10mm; 50,2mm, respectively. 
}
\label{fig:seventeen}
\end{figure}

\begin{figure}
\centerline{ \includegraphics[width=10.cm]{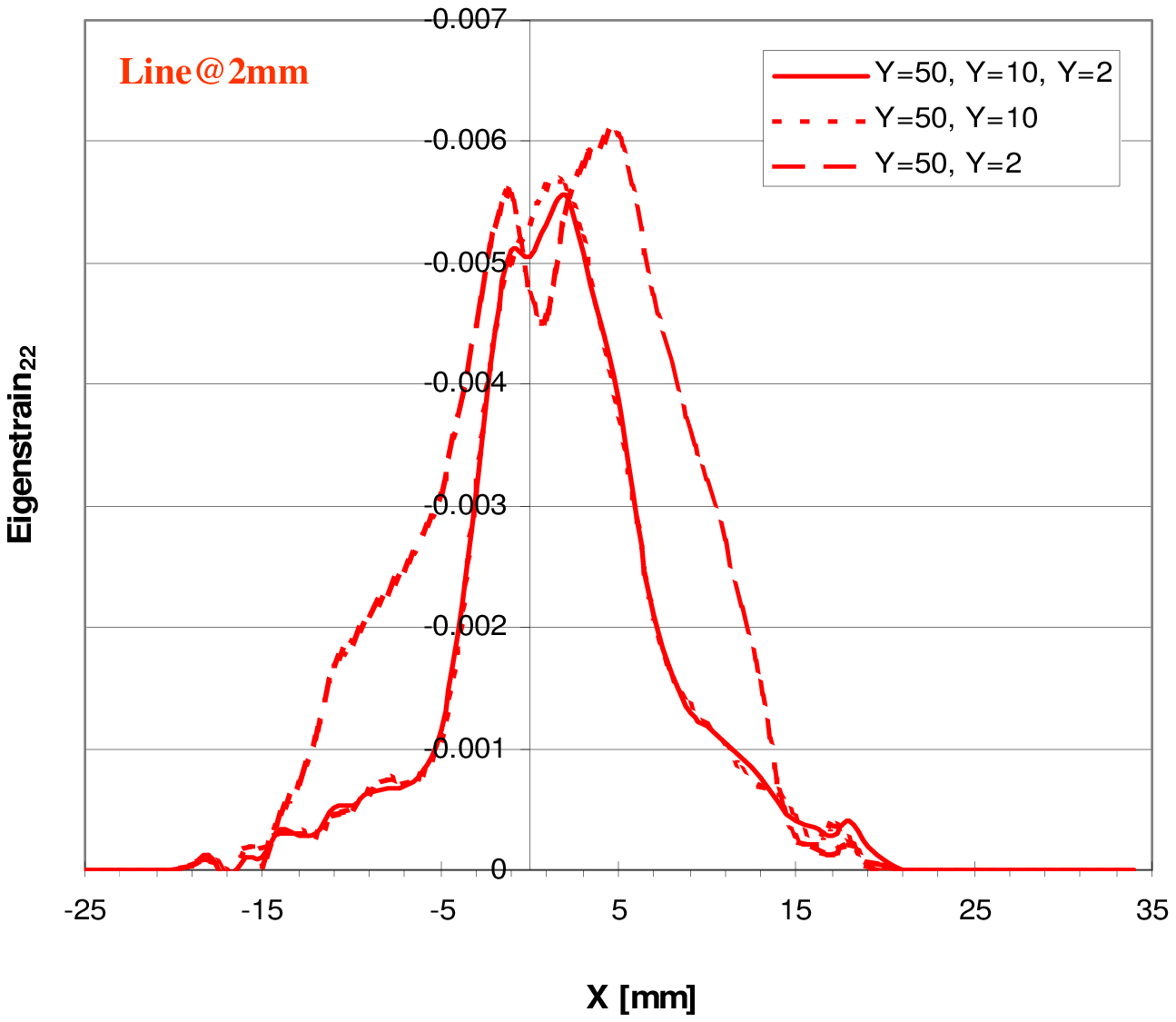} }
\caption{
Comparison plot for the eigenstrain variation perpendicular to the weld line at $y={\rm 2mm}$ obtained from three models using the data for 50,10,2mm; 50,10mm; 50,2mm, respectively. 
}
\label{fig:eighteen}
\end{figure}

\begin{figure}
\centerline{ \includegraphics[width=10.cm]{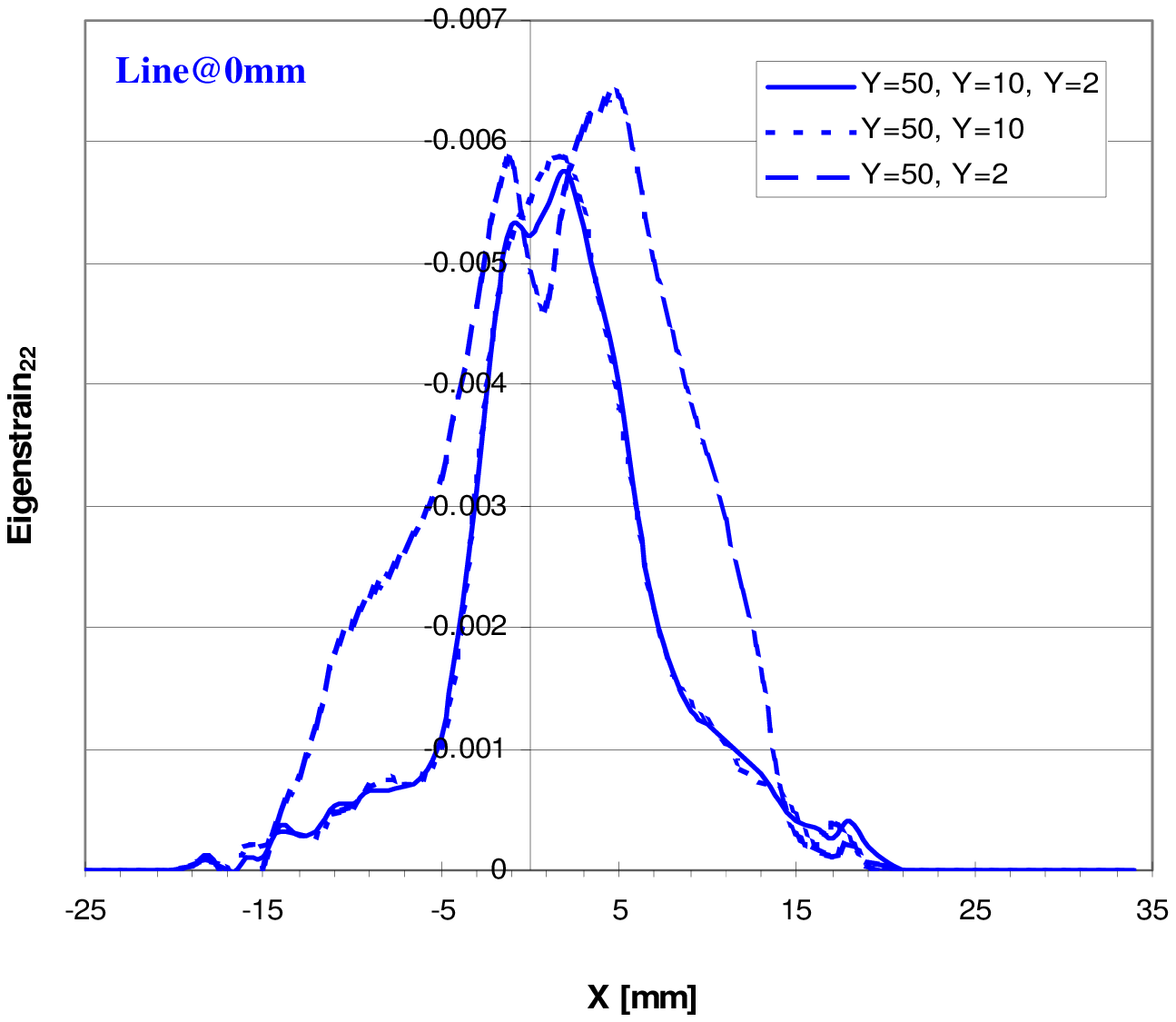} }
\caption{
Comparison plot for the eigenstrain variation perpendicular to the weld line at $y={\rm 50mm}$ (the lower edge of the weld plate) obtained from three models using the data for 50,10,2mm; 50,10mm; 50,2mm, respectively. 
}
\label{fig:nineteen}
\end{figure}

It may appear at first glance that the proposed reconstruction procedure is akin to some kind of a trick, since the amount of information presented in Figures \ref{fig:eight} and \ref{fig:ten} seems significantly greater than that originally avalable from the measurements in Figure \ref{fig:two}. In fact, the reconstructions shown in Figures \ref{fig:eight} and \ref{fig:ten} are not just interpolations, and do contribute additional information to the analysis. By the very nature of the reconstruction process the possible fields included in the consideration are only those that satisfy all the requirements of continuum mechanics. This amounts to a very significant additional constraint being placed on data interpretation. Provided the analysis of the experimental data is carried out in terms of eigenstrain, all of the predicted fields necessarily conform to these constraints, furnishing additional insight into the residual stress field being studied. 

\section{\label{sec:sen} Solution sensitivity to the amount of data available} 

The question that we would like to tackle further concerns the sensitivity of the solution, i.e. the determined eigenstrain distribution and the reconstructedelastic fields, to the selection of experimental data included in the analysis. Instead of attempting to provide a general analytical answer to this question at this point we perform some tests on the data set available within this study, as follows. 

In the results shown in the previous section all of the experimental data available was used in the reconstruction. In the discussion that follows below this model will be referred to as the Y=50,Y=10,Y=2 model, since the data along these lines was used in the reconstruction. We now perform variational eigenstrain analysis on two further models, the Y=50,Y=10 and Y=50,Y=2 models, i.e. omitting line Y=2 and Y=10 respectively.

Figure \ref{fig:twelve} shows that the reconstructed residual stress $\sigma_{22}$ plot along the line at 50mm from the lower edge of the weld plate is quite insensitive to the omission of some data.

Figure \ref{fig:thirteen} represents the plot of $\sigma_{yy}$ along the line 10mm from the lower edge of the weld plate. Clearly the greatest deviation from the complete model results is found in the Y=50,Y=2 model, in which the data along the line Y=10 itself was omitted. However, it s encouraging to note that the qualitative nature of the residual stress distribution is reconstructed correctly, although quantitative difference in the magnitude is observed. Note, however, that this is in fact the result of prediction of the residual stress from measurements conducted remotely from the point of interest made without recourse to any process model whatsoever.

Figure \ref{fig:fourteen} represents the plot of $\sigma_{yy}$ along the line 2mm from the lower edge of the weld plate. Comparison between predictions made by different models once again demonstrates considerable stability of the prediction with respect ot data omission.

Figure \ref{fig:fifteen} presents the plot of $\sigma_{yy}$ along the line $x=0$, i.e. the weld line. It is found that the agreement between the models Y=50,Y=10,Y=2 and Y=50,Y=10 is remarkably good. However, comparison between the Y=50,Y=10,Y=2 and Y=50,Y=2 models shows that omitting Y=10 exerts a significant influence on the predicted residual stress distribution. Note, however, that the data along the $x-$direction is very sparse in the first place, so perhaps this result is not entirely unexpected.

Figure \ref{fig:sixteen} shows the comparison plot for the eigenstrain variation perpendicular to the weld line at $y={\rm 50mm}$ obtained from three models using the data for 50,10,2mm; 50,10mm; 50,2mm, respectively. Remarkable stability of eigenstrain determination with respect to data omission is observed here.

Figure \ref{fig:seventeen} shows the comparison plot for the eigenstrain variation perpendicular to the weld line at $y={\rm 10mm}$ obtained from three models using the data for 50,10,2mm; 50,10mm; 50,2mm, respectively. The two models using the data from the 10mm line show a very close agreement, while the model Y=50,Y=2 shows some deviation, although even in that case the agreeement remains good.

Figure \ref{fig:eighteen} shows the comparison plot for the eigenstrain variation perpendicular to the weld line at $y={\rm 2mm}$ obtained from three models using the data for 50,10,2mm; 50,10mm; 50,2mm, respectively. Once again, agreement between the models remains good evenwhen the data from the line in question is omitted.

Finally, Figure \ref{fig:nineteen} shows the comparison plot for the eigenstrain variation perpendicular to the weld line at $y={\rm 0mm}$ (edge of the plate) obtained from three models using the data for 50,10,2mm; 50,10mm; 50,2mm, respectively, confirming the stability of the eigenstrain determination procedure.

The above analysis does not aim to provide a rigorous proof of the regularity or stability of the proposed inversion procedure. However, it does serve to illustrate that the removal of some data (or its absence in the first place) does not appear to lead to any significant artefacts that might raise doubts in the utility of the proposed approach. 

\section{\label{sec:conc} Conclusion}

It is the authors' belief that the variation approach to eigenstrain determination and residual stress reconstruction introduced in the present paper has the potential to provide very significant improvement in the quality of experimental data interpretation for the purpose of residual stress assessment. The scope of the newly proposed approach is very wide: it can be used with some success to study the data form hole drilling, slitting, Sachs boring and many either destructive and non-destructive technqiues. Furthermore, the eigenstrains introduced into the finite element model in the way described here provide an excellent framework for considering subsequent inelastic deformation mechanism accompanying various processing operations and in situ loading, including creep and/or relaxation during heat treatment, residual stress evolution in fatigue, etc. These research directions are being pursuded by the authors.

\section*{Acknowledgements}

The authors would like to acknowledge the support of UK Department of Trade and Industry and Rolls-Royce plc under project ADAM-DARP.



\begin{thebibliography}{1}

\bibitem{prime}
M.B. Prime, R.J. Sebring, J.M. Edwards, D.J. Hughes, P.J. Webster.
\newblock Laser Surface-Contouring and Spline Data-Smoothing for Residual Stress Measurement. 
\newblock {\em Experimental Mechanics}, 44(2):176-184, 2004.

\bibitem{JSR}
A.M.Korsunsky, S.P.Collins, R.A.Owen, M.R.Daymond, S.Achtioui, and
  K.E.James.
\newblock Fast residual stress mapping using energy dispersive synchrotron
  x-ray diffraction on station 16.3 at the SRS.
\newblock {\em Journal of Synchrotron Radiation}, 9:77--81, 2002.

\bibitem{epsrc}
A.M.Korsunsky et~al.
\newblock Synchrotron stress analysis laboratory (sysal) - a new process and
  design optimisation facility.
\newblock {\em EPSRC Proposal}, 2001.

\bibitem{liu}
J.Liu, K.Kim, M.Golshan, D.Laundy, and A.M.Korsunsky.
\newblock Energy calibration and full-pattern refinement for strain analysis
  using energy-dispersive and monochromatic x-ray diffraction.
\newblock {\em Journal of Applied Crystallography}, 38:661--667, 2005.

\bibitem{mura}
T. Mura.
\newblock {\em Micromechanics of Defects in Solids.}
\newblock Martinus Nijhoff, Dordrecht, 1987.

\bibitem{JSA}
A.M.Korsunsky.
\newblock On the modelling of residual stresses due to surface peening using
  eigenstrain distributions.
\newblock {\em Journal of Strain Analysis for Engineering Design}, 40, 2005.

\bibitem{Brooks}
J.W.Brookes and P.J.Bridges.
\newblock Metallurgical stability of inconel alloy 718.
\newblock {\em Superalloys '88}, pages 33--42, 1988.

\bibitem{Guest}
R.Guest.
\newblock {\em DoITPoMS Micrographs}, 719, 2005.

\end{thebibliography}

\end{document}